\documentclass[preprint]{emulateapj} 
\begin{document}
 
\title{Dissipative Divergence of Resonant Orbits}  
\author{Konstantin Batygin$^1$ \& Alessandro Morbidelli$^2$} 

\affil{$^1$Division of Geological and Planetary Sciences, California Institute of Technology, Pasadena, CA 91125} 
\affil{$^2$Departement Cassiop$\mathrm{\acute{e}}$e: Universite de Nice-Sophia Antipolis, Observatoire de la C$\mathrm{\hat{o}}$te dÕAzur, 06304 Nice, France}
\email{kbatygin@gps.caltech.edu}

\begin{abstract} 

A considerable fraction of multi-planet systems discovered by the observational surveys of extrasolar planets reside in mild proximity to first-order mean motion resonances. However, the relative remoteness of such systems from nominal resonant period ratios (e.g. 2:1, 3:2, 4:3) has been interpreted as evidence for lack of resonant interactions. Here we show that a slow divergence away from exact commensurability is a natural outcome of dissipative evolution and demonstrate that libration of critical angles can be maintained tens of percent away from nominal resonance. We construct an analytical theory for the long-term dynamical evolution of dissipated resonant planetary pairs and confirm our calculations numerically. Collectively, our results suggest that a significant fraction of the near-commensurate extrasolar planets are in fact resonant and have undergone significant dissipative evolution.

\end{abstract} 

\maketitle

\section{Introduction}

Among the most unexpected discoveries brought forth by extrasolar
planetary surveys to date has been the identification of numerous
planetary bodies that reside in close proximity to their host
stars. Planets of this sort are of great scientific interest because
they represent a class of objects unavailable for study in our own
solar system. In turn, observational characterization of such
planetary systems can yield avenues towards identifying specific
physical/dynamical behavior that does not occur locally, thus broadening
our knowledge of the possible evolutions of planetary systems.

A readily apparent dynamical feature of close-in extra-solar planetary systems, highlighted by observational surveys such as the \textit{Kepler} mission \citep{2011ApJ...728..117B, 2011ApJS..197....8L}, is the prominence of near mean-motion commensurabilities (i.e. integer period ratios) among sub-giant planets (Figure 1). Accordingly, understanding how close-in planetary systems attain near-resonant orbital architectures is the primary focus of this work.

The process of resonant locking requires slow, convergent orbital evolution of planetary bodies \citep{1965MNRAS.130..159G, 1976ARA&A..14..215P}. It is likely that torques associated with disk-driven migration often lead to resonant coupling, and it has been suggested that near-exact commensurability should be maintained as planets travel through their proto-planetary disks \citep{2007ApJ...654.1110T, 2008A&A...482..677C}. However, the onset of magneto-rotational instability \citep{1991ApJ...376..214B} and the associated turbulence in protoplanetary disks can act to disrupt mean-motion resonances \citep{2008ApJ...683.1117A, 2009A&A...497..595R, 2011ApJ...726...53K}. Thus, if disks are violently turbulent, resonant objects should be rare. 

As already hinted above, the observations show that there exists a characteristic regime in between the two extremes, and the precise dynamical nature of this regime is elusive. Particularly, planets often reside sufficiently far away (a few percent or more) from their nominal first-order resonant locations (i.e. period ratios of 2:1, 3:2, 4:3) to be readily interpreted as non-resonant. Yet the preference for orbits just wide of resonance and a characteristic pile-up of near-resonant objects (Fig. 1) is suggestive of a common evolutionary path. Indeed, the mechanism responsible for such configurations has been noted to be a subject of great theoretical interest \citep{2012arXiv1202.6328F}.

It is possible in principle that most sub-giant planets arrive onto their close-in orbits in resonance and subsequently diverge away from exact commensurability due to tidal dissipation. Tides alone affect the semi-major axes only on very long timescales (often much longer than the Hubble time). However,  as shown by the non-linear perturbative calculations and $N$-body simulations aimed at reproducing the orbital configurations of the HD40307 \citep{2010MNRAS.405..573P} as well as GL581 and HD10180 \citep{2011CeMDA.111...83P} systems, resonant interactions can be quite effective at converting tidal eccentricity damping (which acts much faster) into a divergence of the orbital semi-major axes of the resonant bodies.  In particular, the said simulations suggest that resonant coupling can be maintained far from nominal resonant locations and significantly aids in enhancing orbital divergence.

\begin{figure}
\label{keplerdata}
\includegraphics[width=0.5\textwidth]{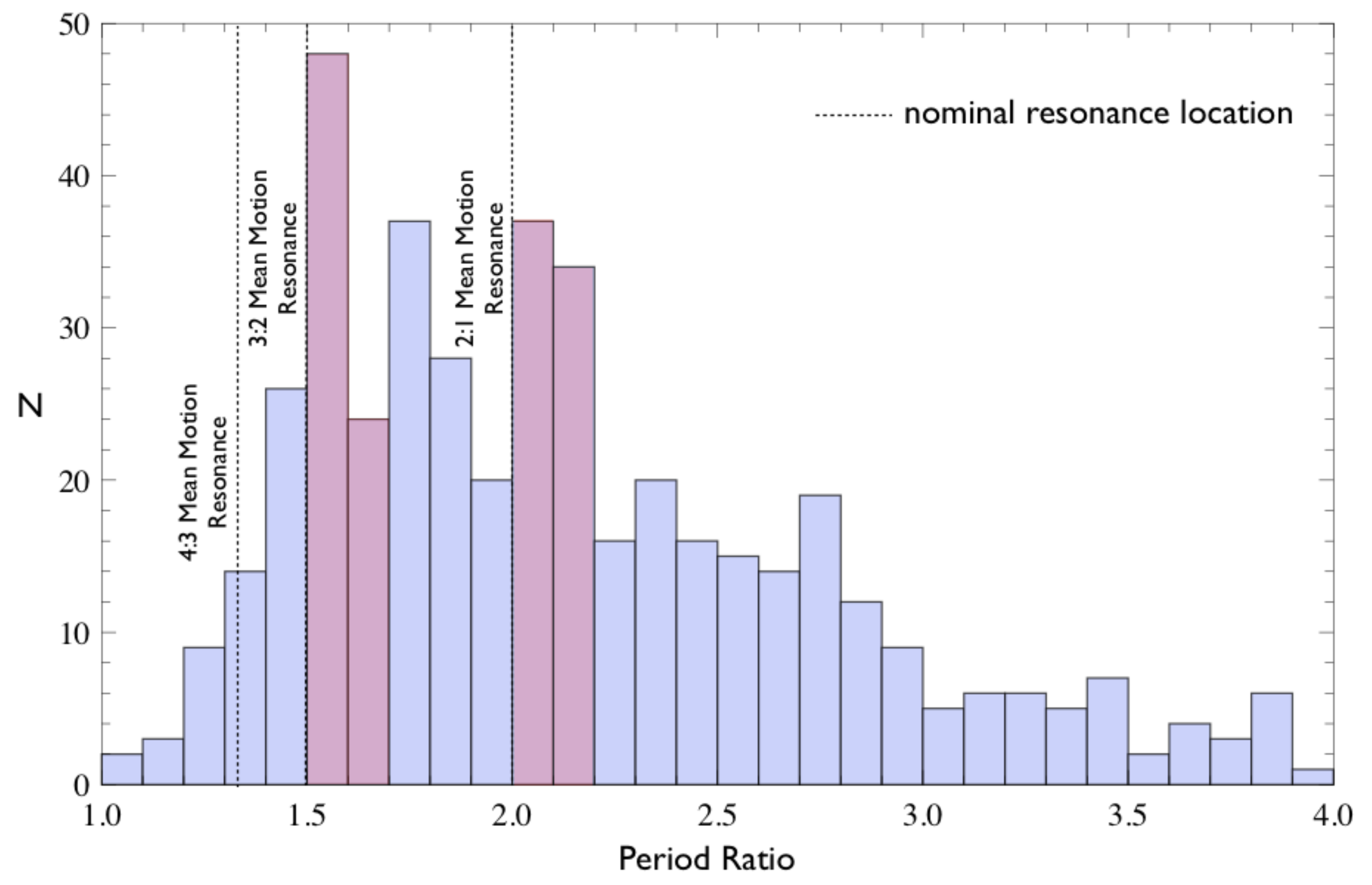}
\caption{A histogram of the period ratios of all planet pairs detected by the $Kepler$ mission with no filters on planetary radius or orbital period (http://planetquest.jpl.nasa.gov/kepler). In systems where more than two planets are present, only the neighboring period ratios are reported. Note the highlighted enhancement of objects immediately outside of the common (2:1 and 3:2) first-order mean motion resonances. }
\end{figure}

The calculations performed by \citep{2010MNRAS.405..573P} motivate our development of a general qualitative understanding of the orbital evolution of close-in resonant planetary systems subject to dissipative effects. Thus, the development of an analytical theory for dissipative divergence of resonant orbits is the primary focus of this paper. The number of well-characterized systems within the $Kepler$ sample remains limited and estimation of planetary masses from radii alone is generally risky \citep{1982AREPS..10..257S, 2011ApJ...738...59R}. Consequently, in this work, we shall concentrate our efforts on characterization of the physical process rather than reproduction of any particular orbital architecture. Still, we argue that the interplay between resonant effects and tidal dissipation is the primary mechanism by which planets attain near-commensurate orbits. \citet{2012arXiv1204.2555L} arrived at many of the results presented in this work simultaneously and independently; their paper was posted on arxiv.org at the same time as this one. 

The paper is organized as follows. In section 2, we set the stage by developing an integrable approximation to the conservative dynamics of a resonant pair at low eccentricities and validate the theory by comparison with $N$-body simulations. In section 3, we introduce dissipation into the problem and show that tidal effects drive the system towards a quasi-stationary state that is characterized by an irreversible drift away from nominal resonance, where the inner planet's orbit decays at a rate that is faster than that expected from the direct tidal effect, while the outer planet gains orbital energy. In section 4, we discuss the extension of our formalism to multi-resonant systems. Subsequently, we conclude and discuss our results in section 5. 
 
\section{Conservative Dynamics of a Resonant Planetary Pair}

Resonant dynamics of planetary pairs have been studied by numerous authors in the past (see Ch.8 of \citet{1999ssd..book.....M} and the references therein). This work builds on their contributions. 

Our eventual goal is to construct an analytical model for the long-term evolution of resonant orbits under dissipative effects. Before complicating the picture with dissipation, however, we must first build a purely analytical model for conservative resonant interactions. Thus, in this section, we shall derive a simple, physically intuitive closed-form solution for the time-evolution of a resonant planetary pair. Accordingly, we shall first work in the spirit of classical perturbation theory (e.g. \citet{1966IAUS...25..197M, 1986sate.conf..159P}) and employ numerical calculations primarily as a means of confirmation.

Let us begin by considering a quasi-integrable Hamiltonian of the form
\begin{equation}
\label{H}
\mathcal{H} = \mathcal{H}_{\rm{kep}} + \mathcal{H}_{\rm{res}} + \mathcal{O} (e^2 , i^{2}) ,
\end{equation}
where
\begin{equation}
\mathcal{H}_{\rm{kep}} = -G \frac{M m_1}{2 a_1} -G \frac{M m_2}{2 a_2}
\end{equation}
is the Keplerian Hamiltonian and 
\begin{eqnarray}
\mathcal{H}_{\rm{res}} = &-&G \frac{m_1 m_2}{a_2} ( f_{\rm{res}}^{(1)} e_1 \cos(k \lambda_2 - (k-1) \lambda_1 - \varpi_1)  \nonumber \\
&+& f_{\rm{res}}^{(2)} e_2 \cos(k \lambda_2 - (k-1) \lambda_1 - \varpi_2) )
\end{eqnarray}
is the first-order $k: k-1, k \in \mathbb{Z}$ resonant perturbation. Here, the orbital elements take on their standard notation, $M$ is the mass of the central star and $m_1, m_2$ are the masses of the planets with the subscript $1$ and $2$ referring to the inner and outer planets respectively. The quantities $f_{\rm{res}}^{(1)}$ and $f_{\rm{res}}^{(2)}$ depend on the semi-major axis ratio $(a_1/a_2)$ only and are tabulated in the literature (see for example \citet{1999ssd..book.....M}).

Because Keplerian orbital elements are not canonically conjugated, we revert to Poincar$\acute{\rm{e}}$ variables for further calculations:
\begin{eqnarray}
\Lambda &=& m \sqrt{G M a}, \ \ \ \ \ \ \ \ \ \ \ \ \ \ \ \ \ \ \ \ \lambda = \mathcal{N} + \varpi \\
\Gamma &=& \Lambda (1 - \sqrt{1-e^2}) \approx \Lambda \ e^2/2, \ \ \ \gamma = - \varpi,
\end{eqnarray}
where $\mathcal{N}$ is the mean anomaly and the indexe $1,2$ are omitted for simplicity. In terms of the Poincar$\acute{\rm{e}}$ variables, the Hamiltonians, $\mathcal{H}_{\rm{kep}}$ and $\mathcal{H}_{\rm{res}}$ read:
\begin{equation}
\label{Hkeplambda}
\mathcal{H}_{\rm{kep}} = - \frac{G^2 M^2 m_1^3}{2 \Lambda_1^2} - \frac{G^2 M^2 m_2^3}{2 \Lambda_2^2},
\end{equation}
\begin{eqnarray}
\label{Hrespoincare}
\mathcal{H}_{\rm{res}} &=& - \frac{G^2 M m_1 m_2^3}{\Lambda_2^2} ( f_{\rm{res}}^{(1)} \sqrt{\frac{2 \Gamma_1}{\Lambda_{1}}} \cos(k \lambda_2 - (k-1) \lambda_1 + \gamma_1)  \nonumber \\
&+& f_{\rm{res}}^{(2)}  \sqrt{\frac{2 \Gamma_2}{\Lambda_{2}}} \cos(k \lambda_2 - (k-1) \lambda_1 + \gamma_2) ).
\end{eqnarray}

\begin{figure*}
\label{21MMRcons}
\includegraphics[width=1\textwidth]{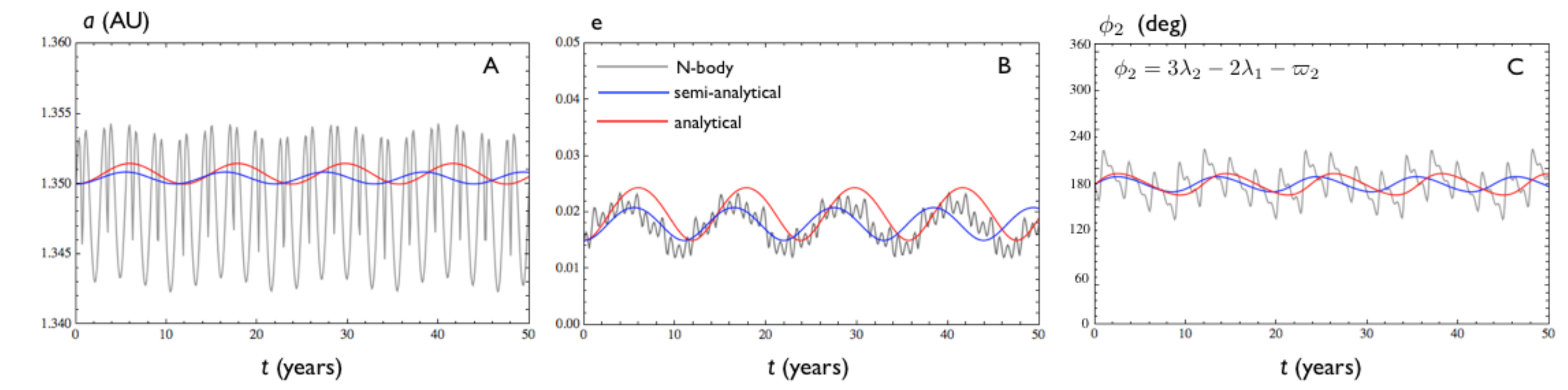}
\caption{Orbital evolution of a nearly mass-less ($m = 10^{-10} M_{\odot}$) particle in an interior 2:1 mean motion resonance with a Jupieter-mass object ($m = 10^{-3} M_{\odot}$) with a semi-major axis of $a = 1$ AU. The evolution is shown over 50 orbital periods of the perturbing object, corresponding to $\sim 5$ resonant cycles. The panels A, B, and C show the variation in the particle's semi-major axes, eccentricity and the critical resonant angle respectively. The red curve was obtained analytically utilizing the framework developed in section 2. The blue curve was obtained by numerically integrating the equations of motion that arise from the Hamiltonians (\ref{Hkeplambda}) and (\ref{Hrespoincare}). The gray curve is a result of a direct N-body simulation.}
\end{figure*}

As already implied by equation ($\ref{H}$), we shall work to first order in eccentricity, neglecting secular effects and resonances of order greater than unity. Generally, $\mathcal{H}$ only constitutes a good approximation to the true dynamics of a planetary pair in the vicinity of a mean-motion resonance. 

Because the perturbation $\mathcal{H}_{\rm{res}}$ is of order $e$, we expect that the semi major axes can change by $\mathcal{O}(\sqrt{e})$ relative to their nominal, resonant values. Thus, we expand the terms in $\mathcal{H}_{\rm{kep}}$ to second order in $\delta \Lambda = \Lambda - [\Lambda]$, where $[\Lambda]$ is the nominal value of $\Lambda$:
\begin{eqnarray}
\mathcal{H}_{\rm{kep}} = &-& \frac{G^2 M^2 m_1^3 }{2 [\Lambda]_1^2} + \frac{G^2 M^2 m_1^3}{[\Lambda]_1^3}\delta \Lambda_1 - \frac{3 G^2 M^2 m_1^3}{2 [\Lambda]_1^4}\delta \Lambda_1^2 \nonumber \\
&-&\frac{G^2 M^2 m_2^3 }{2 [\Lambda]_2^2} + \frac{G^2 M^2 m_2^3}{[\Lambda]_2^3}\delta \Lambda_2 - \frac{3 G^2 M^2 m_2^3}{2 [\Lambda]_2^4}\delta \Lambda_2^2 \nonumber \\
&+& \mathcal{O}(\delta \Lambda_1^3, \delta \Lambda_2^3).
\end{eqnarray}
Consistently, we evaluate $\mathcal{H}_{\rm{res}}$ in (6) at $
[\Lambda]$, as it is already of order $\mathcal{O}(e)$. Constant terms are
dynamically unimportant and can thus be dropped from the Hamiltonian,
implying $\delta \Lambda \rightarrow \Lambda$ and $\delta \Lambda^2
\rightarrow \Lambda^2 - 2 \Lambda [\Lambda]$:
\begin{eqnarray}
\mathcal{H}_{\rm{kep}} &=& \frac{4 G^2 M^2 m_1^3 \Lambda_1}{[\Lambda_1]^3} + \frac{4 G^2 M^2 m_2^3 \Lambda_2}{[\Lambda_2]^3} \nonumber \\
 &-& \frac{3 G^2 M^2 m_1^3 \Lambda_1^2}{2 [ \Lambda_1]^4} - \frac{3 G^2 M^2 m_2^3 \Lambda_2^2}{2 [ \Lambda_2]^4}.
\end{eqnarray}
Note that the planetary mean motion is given by
\begin{equation}
n =\frac{d \lambda}{dt} = \frac{\partial \mathcal{H}_{\rm{kep}}}{ \partial \Lambda} = \frac{G^2 M^2 m^3 }{\Lambda^3}.
\end{equation}
As a result, $\mathcal{H}_{\rm{kep}}$ can be rewritten in a compact form:
\begin{equation}
\mathcal{H}_{\rm{kep}} = 4([n]_1 \Lambda_1 + [n]_2 \Lambda_2) - \frac{3}{2}([h]_1 \Lambda_1^2 +  [h]_2 \Lambda_2^2),
\end{equation}
where $[h] = [n]/[\Lambda] = m/[a]^2$.

Although $\mathcal{H}_{\rm{kep}}$ is now expressed in a simple form, $H_{\rm{res}}$ remains cumbersome largely due to the formulation of the resonant angles which appear as cosine arguments. Let us employ a canonical transformation of coordinates, utilizing the following generating function of the second kind:
\begin{eqnarray}
F_2 &=& \lambda_1 \Psi_1 + \lambda_2 \Psi_2 + (k \lambda_2 - (k-1) \lambda_1 + \gamma_1)\Phi_1 \nonumber \\
&+& (k \lambda_2 - (k-1) \lambda_1 + \gamma_2)\Phi_2,
\end{eqnarray}
where $\Psi$ and $\Phi$ are new momenta. Upon application of the transformation equations
\begin{equation}
\Lambda = \frac{\partial F}{\partial \lambda} \ \ \ \ \ \Gamma = \frac{\partial F}{\partial \gamma}
\end{equation}
we obtain new canonically conjugated action-angle variables
\begin{eqnarray}
\Psi_1 &=& \Lambda_1 + (k -1) (\Phi_1 + \Phi_2)  \ \ \ \psi_1 = \lambda_1 \nonumber \\
\Psi_2 &=& \Lambda_2 - k (\Phi_1 + \Phi_2)  \ \ \ \ \ \ \ \ \ \ \  \psi_2 = \lambda_2 \nonumber \\
\Phi_1 &=& \Gamma_1 \ \ \ \ \ \ \phi_1 = k \lambda_2 - (k-1) \lambda_1 + \gamma_1   \nonumber \\
\Phi_2 &=& \Gamma_2 \ \ \ \ \ \  \phi_2 = k \lambda_2 - (k-1) \lambda_1 + \gamma_2.
\end{eqnarray}
In terms of these variables, the resonant contribution to $\mathcal{H}$ is expressed as follows:
\begin{eqnarray}
\label{sfmr_res}
\mathcal{H}_{\rm{res}} &=& - \frac{G^2 M m_1 m_2^3}{[\Lambda]_2^2} ( f_{\rm{res}}^{(1)} \sqrt{\frac{2 \Phi_1}{[\Lambda]_{1}}} \cos(\phi_1)  \nonumber \\
&+& f_{\rm{res}}^2  \sqrt{\frac{2 \Phi_2}{[\Lambda]_{2}}} \cos(\phi_2) ).
\end{eqnarray}
while the Keplerian contribution reads:
\begin{eqnarray}
\label{sfmr_kep}
\mathcal{H}_{\rm{kep}} &=& 4[n]_1 (\Psi_1 - (k-1)(\Phi_1 + \Phi_2) ) \nonumber \\
& + & 4 [n]_2 (\Psi_2 + k (\Phi_1 + \Phi_2) )  \nonumber \\
&-& \frac{3}{2}[h]_1 (\Psi_1 - (k-1)(\Phi_1 + \Phi_2) )^2  \nonumber \\
&-& \frac{3}{2}[h]_2 (\Psi_2 + k (\Phi_1 + \Phi_2)^2 ).
\end{eqnarray}
The transformation to new variables allows us to make further simplifications to $\mathcal{H}_{\rm{kep}}$. Specifically, because $\partial{H}/\partial{\psi} = 0$, $\Psi_1$ and $\Psi_2$ are constants of motion, allowing us to drop additional terms. It is further instructive to recall that $\Phi \propto e^2$. Consequently, if $e \ll 1$, non-linear terms proportional to  $\Phi_1^2$, $\Phi_2^2$, and $\Phi_1\Phi_2$ can be neglected. This approximation filters out chaotic dynamics from the Hamiltonian and therefore will not yield an adequate representation of the evolution of the system in the resonances overlap region \citep{1979PhR....52..263C, 1980AJ.....85.1122W}. However as will be shown below, this assumption is well satisfied in the calculations of interest. Upon making these simplifications, the Keplerian Hamiltonian is simply
\begin{eqnarray}
\mathcal{H}_{\rm{kep}} &=& (4(k [n]_2 - (k-1)[n]_1)) \nonumber \\
&+& 3 ([h]_1 (k - 1) \Psi_1 - [h]_2 k \Psi_2 ))(\Phi_1 + \Phi_2).
\end{eqnarray}
Note that by definition, $(k [n]_2 - (k-1)[n]_1) = 0$ because it signifies exact resonance. As a result, only terms proportional to $[h]$ remain in $\mathcal{H}_{\rm{kep}}$.

\begin{figure*}
\label{32MMRcons}
\includegraphics[width=1\textwidth]{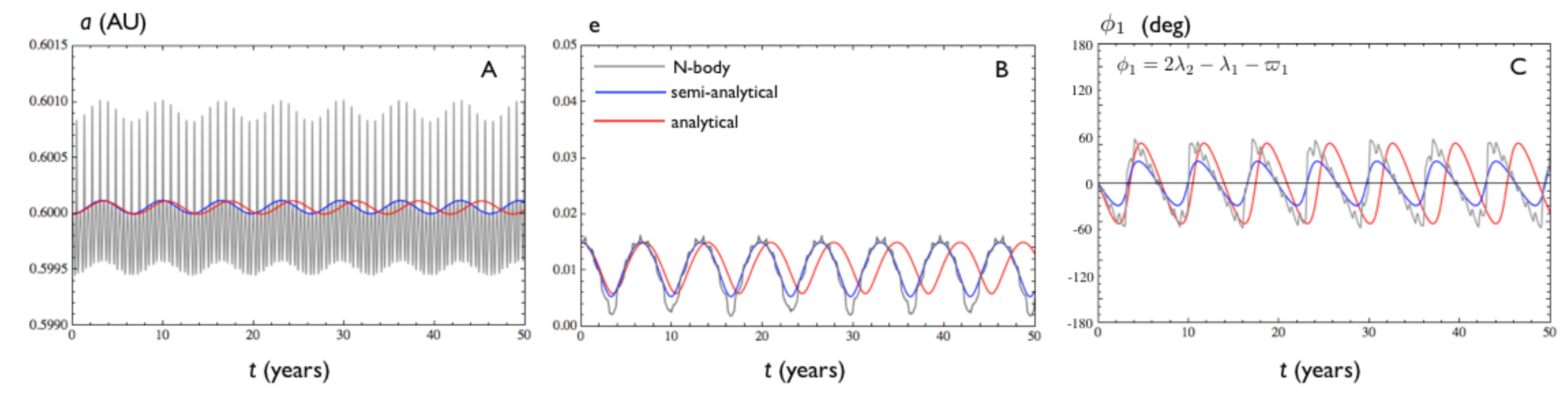}
\caption{Orbital evolution of a nearly mass-less ($m = 10^{-10} M_{\odot}$) particle in an exterior 3:2 mean motion resonance with a Jupieter-mass object ($m = 10^{-3} M_{\odot}$) with a semi-major axis of $a = 1$ AU. The evolution is shown over 50 orbital periods of the perturbing object, corresponding to $\sim 7$ resonant cycles. As in Figure (2), the panels A, B, and C show the variation in the particle's semi-major axes, eccentricity and the critical resonant angle respectively. The red curve was obtained analytically utilizing the framework developed in section 2. The blue curve was obtained by numerically integrating the equations of motion that arise from the Hamiltonians (\ref{Hkeplambda}) and (\ref{Hrespoincare}). The gray curve is a result of a direct N-body simulation.}
\end{figure*}

The full Hamiltonian now takes on a very simple form:
\begin{equation}
\label{Hsimple}
\mathcal{H} = \eta (\Phi_1 + \Phi_2) + \alpha \sqrt{2 \Phi_1} \cos(\phi_1) + \beta \sqrt{2 \Phi_2} \cos(\phi_2),
\end{equation}
where 
\begin{equation}
\eta = 3 ([h]_1 (k - 1) \Psi_1 - [h]_2 k \Psi_2 )
\end{equation}
is related the circulation frequency of the critical angles in an unperturbed case ($m_1=m_2=0$) and is thus a measure of proximity of the planetary pair to exact Keplerian resonance (note that $\eta \rightarrow 0$ as $\Lambda \rightarrow [\Lambda]$ and $\Phi \rightarrow 0$, corresponding to $\Psi = [\Lambda]$) while 
\begin{eqnarray}
\label{alpha}
\alpha &=& - \frac{G^2 M m_1 m_2^3}{[\Lambda]_2^2} \frac{ f_{\rm{res}}^{(1)}}{\sqrt{[\Lambda]_{1}}} \nonumber \\
\beta &=& - \frac{G^2 M m_1 m_2^3}{[\Lambda]_2^2} \frac{ f_{\rm{res}}^{(2)}}{\sqrt{[\Lambda]_{2}}}
\end{eqnarray}
are the strengths of the resonances. 
It is noteworthy that the Hamiltonian (\ref{Hsimple}) represents two decoupled Hamiltonians, each of which has a form similar of the ``second fundamental model of resonance'' \citep{1983CeMec..30..197H}, apart from the missing term, proportional to $\Phi^2$, that we have neglected. 

In the coordinates used up to now, the equations of motion are singular at $\Phi = 0$. However, this singularity can be overcome by switching to mixed cartesian coordinates
\begin{eqnarray}
x = \sqrt{2 \Phi} \sin(\phi) \ \ \ \ \ y = \sqrt{2 \Phi} \cos(\phi)
\end{eqnarray}
via a contact transformation (here, $x$ is identified as the coordinate and $y$ as the momentum). In these coordinates, the Hamiltonian reads
\begin{equation}
\mathcal{H} = \frac{\eta}{2} (x_1^2 + y_1^2 + x_2^2 + y_2^2) + \alpha y_1 + \beta y_2.
\end{equation}
Accordingly, the equations of motion are:
\begin{eqnarray}
\frac{dx_1}{dt} &=& \frac{\partial \mathcal{H} }{\partial y_1} = \alpha + \eta y_1 \nonumber \\
\frac{dx_2}{dt} &=& \frac{\partial \mathcal{H} }{\partial y_2} = \beta + \eta y_2 \nonumber \\
\frac{dy_1}{dt} &=& -\frac{\partial \mathcal{H} }{\partial x_1} = - \eta x_1 \nonumber \\
\frac{dy_2}{dt} &=& -\frac{\partial \mathcal{H} }{\partial x_2} = - \eta x_2.
\end{eqnarray}
Although we can continue to work in terms of the mixed cartesian coordinates, the equations of motion can be re-written in a more compact form by treating $x$ and $y$ as imaginary and real components of a single complex variable 
\begin{equation}
z = \imath x + y.
\end{equation}
Now, the equations of motion can be written down concisely:
\begin{eqnarray}
\label{eqsmotion}
\frac{dz_1}{dt} &=& \imath \alpha + \imath \eta z_1 \nonumber \\
\frac{dz_2}{dt} &=& \imath \beta + \imath \eta z_1,
\end{eqnarray}
and admit the analytical solutions 
\begin{eqnarray}
\label{soln}
z_1 &=& -\frac{\alpha}{\eta} +\mathcal{C}_1 \exp(\imath \eta t) \nonumber \\
z_2 &=& -\frac{\beta}{\eta} +\mathcal{C}_2 \exp(\imath \eta t),
\end{eqnarray}
where $\mathcal{C}_1$ and $\mathcal{C}_2$ are (possibly complex) constants of integration. Note that except for a dependence of the leading term on $z$, equations (\ref{eqsmotion}) are analogous to the complex formulation of the Laplace-Lagrange theory for secular interactions \citep{2002ApJ...564.1024W,  2011ApJ...730...95B}, although the variables take on a different meaning.

\begin{figure*}
\label{21MMRdissipative}
\includegraphics[width=1\textwidth]{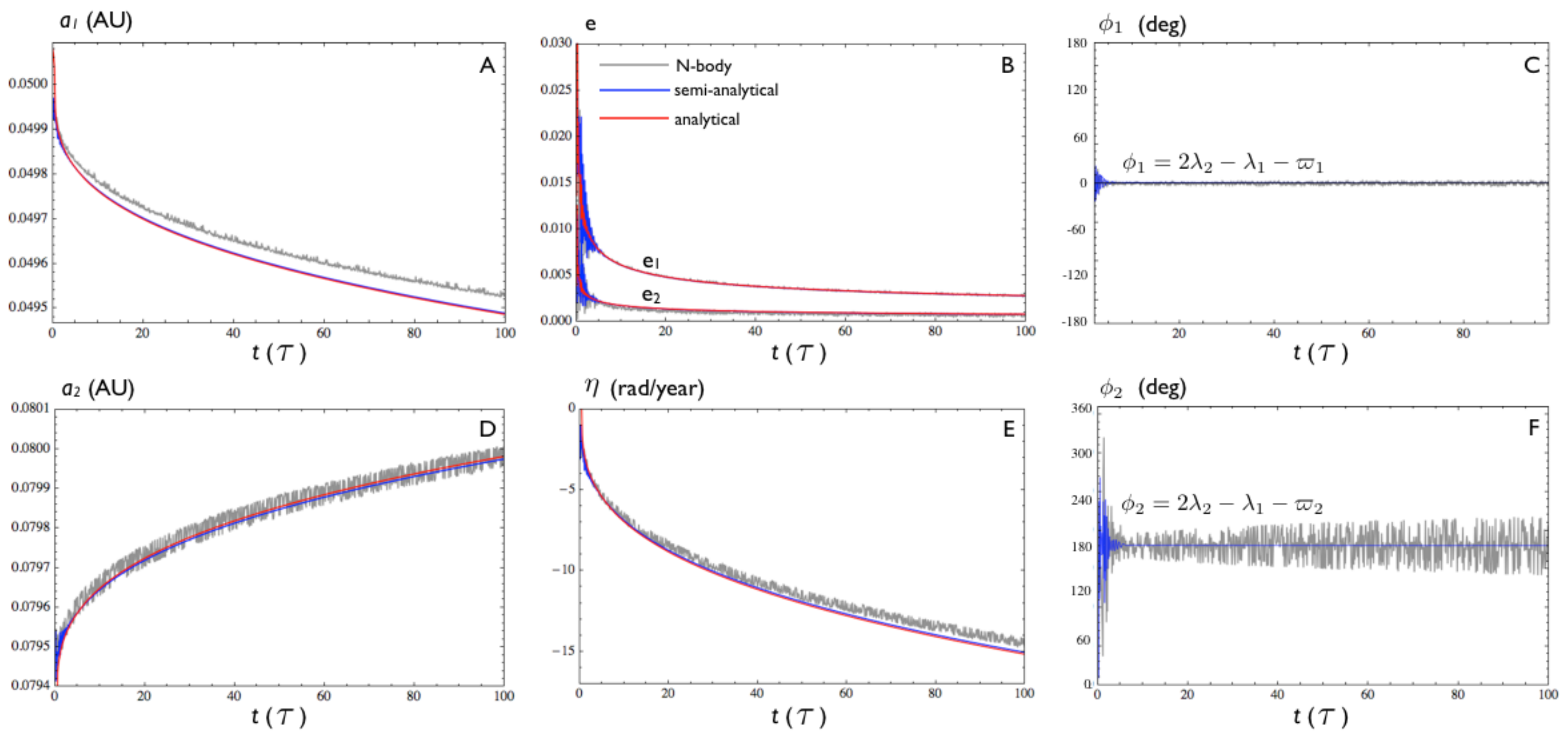}
\caption{Dissipative evolution of an equal-mass ($m_1 = m_2 = 10^{-4}M$) planetary pair in a 2:1 mean motion resonance over $t/\tau = 100$ circularization timescales. Panels A and D show the evolution of the planetary semi-major axes. Note that at all times dissipative interactions give rise to a monotonic divergence of the orbits. This can be further inferred from panel E which shows the measure of proximity to exact resonance, $\eta < 0$ monotonically decreasing. Panels C and F show the evolution of the critical angles. Note that the system attains a state of quasi-equilibrium after $t \sim 5 \tau$. Accordingly, the eccentricity evolution also becomes quasi-stationary after the critical angles collapse to a near-focal state. The red curves were obtained analytically utilizing the framework developed in section 3. The blue curves were obtained by numerically integrating the equations of motion that arise from the Hamiltonians (\ref{Hkeplambda}) and (\ref{Hrespoincare}), augmented with a simple parameterization of tidal dissipation (i.e. equations (\ref{dedttide}) and (\ref{dadttidal})). The gray curves were computed numerically with a direct N-body simulation where dissipation has been taken into account using the tidal framework of \citep{1998ApJ...499..853E}.}
\end{figure*}

Within the context of this model, variations in semi-major axes can be derived from the fact that $\Psi$ remain constants of motion. Examples of the application of the theory are presented in Figures (2) and (3). In both of the illustrated cases, a nearly mass-less ($m = 10^{-10} M_{\odot}$) particle is perturbed by a Jupieter-mass object ($m = 10^{-3} M_{\odot}$) with a semi-major axis of $a = 1$ AU. Figure (2) shows an interior 2:1 mean motion resonance while Figure (3) shows an exterior 3:2 mean motion resonance. The red curves denote analytical theory, the blue curves represent a numerical integration of the non-linear perturbative Hamiltonians (\ref{Hkeplambda}) and (\ref{Hrespoincare}), and the gray curves are the results of numerical $N$-body simulations, performed using the hybrid algorithm of the orbital integration software package \textit{mercury6} \citep{1999MNRAS.304..793C}.  Note that as a consequence of the simplifications made in order to express the analytical solution in closed form, the blue (non-linear perturbative) curve has slightly different frequency and amplitude of oscillation relative to the red (analytical) curve, although the two curves exhibit the same qualitative behavior. However, in addition to the resonant variations, the grey ($N$-body) curve shows non-resonant, short-period oscillations, that are filtered out by retaining only the resonant terms in the Hamiltonian. These short-periodic oscillations are unimportant to the problem at hand, as they do not contribute to the time-averages of the resonant angles. Note also that, although the particles in both examples are relatively far away from nominal resonance, the critical angles remain in libration. 

\section{Dissipative Dynamics of a Resonant Planetary Pair}
There exists an abundance of circumstances where the evolution of a planetary system cannot be described in terms of strictly conservative interactions. For example, planets embedded in protoplanetary disks experience dissipative forces exerted by the gaseous nebula \citep{2002ApJ...567..596L}, while planets that reside on orbits that are in close proximity to their host stars are subject to tidal friction \citep{2001ApJ...548..466B} (in this work, we shall concentrate on the latter). In the extrasolar context, tidal dissipation usually results in the decay of orbital eccentricity and semi-major axes. 

With the exception of special configurations, the characteristic timescales for the decay of eccentricity and semi-major axes differ significantly (often by orders of magnitude). This is in part because the changes in eccentricity are controlled by the rate of angular momentum exchange in the system, while changes in the semi-major axes are largely governed by the rate of energy dissipation, which is usually a much slower process. As a result for the purposes of this work, we shall invoke separation of timescales and treat the decays of $e$ and $a$ independently. 
For $e \ll 1$, the orbit-averaged rate of tidal eccentricity decay is given by \citep{1966Icar....5..375G}:
\begin{equation}
\label{dedttide}
\left(\frac{d e}{dt}\right)_{tide}  = -e \frac{21 [n]}{2} \frac{k}{Q} \frac{M}{m} \left(\frac{R}{[a]}\right)^{5} = - \frac{e}{\tau_e},
\end{equation}
where $k$ is the planetary Love number, $Q$ is the tidal quality factor (note that dissipation within the host-star is neglected as usual), and $R$ is the planetary radius. Noting that $|z| \simeq e \sqrt{[\Lambda]}$, it is trivial to incorporate eccentricity decay into equations (\ref{eqsmotion}):
\begin{eqnarray}
\label{eqsmotiondiss}
\frac{dz_1}{dt} &=& \imath \alpha + \imath \eta z_1 - \frac{z_1}{\tau_{e_1}} \nonumber \\
\frac{dz_2}{dt} &=& \imath \beta + \imath \eta z_1  - \frac{z_2}{\tau_{e_2}}.
\end{eqnarray}
Since the equations of motion remain linear in $z$, they admit solutions that 
are formally similar to (\ref{soln}):
\begin{eqnarray}
\label{solndiss}
z_1 &=& -\frac{\alpha}{\eta+\imath/\tau_{e_1}} + \mathcal{C}_1 \exp(\imath \eta t - t/\tau_{e_1}) \nonumber \\
z_2 &=& -\frac{\beta}{\eta+\imath/\tau_{e_2}} + \mathcal{C}_2 \exp(\imath \eta t - t/\tau_{e_2}).
\end{eqnarray}
Note that the eccentricity damping timescale of the second body in the equation above is $\tau_{e_2}$. Depending on $Q$, this timescale can appear to greatly exceed $\tau_{e_1}$. However, it is important to keep in mind that in reality, variations in $\phi_1$ and $\phi_2$ are coupled because both give rise to changes in the planetary semi-major axes. This means that tidal dissipation of the inner planet's eccentricity also damps the outer planet's eccentricity resonantly. Furthermore, the first and the second planet are also coupled through a secular term of the form $\mathcal{H}_{\rm{sec}} \propto e_1 e_2 \cos(\varpi_1-\varpi_2)$, that we have neglected in the Hamiltonian. Through this secular interaction, tidal damping on $e_1$ is translated to $e_2$ as well (albeit on a longer timescale), even if there is no direct damping on $e_2$ (i.e. $ \tau_{e_2}=\infty$; \citet{2002ApJ...564.1024W, 2007MNRAS.382.1768M}). 

Because the dissipation is applied directly on the actions, Hamiltonian properties of the solution such as the conservation of phase-space area bounded by the orbit are destroyed. On a timescale of a $\sim$ few $\tau_e$, the second terms in the solutions (\ref{solndiss}) will decay away, making the phase-space area bounded by the orbit tend to zero. This has a number of important physical implications. First of all, this removes the dependence of the long term ($t \gg \tau_z$) solution on the initial conditions. Second, the fact that explicit time-dependence of the solution is also lost, suggests that the eccentricity dynamics falls onto a fixed point attractor, characterized by constant actions (i.e. eccentricities) and angles \citep{2011CeMDA.111..219B}. Specifically, assuming that $1/\tau_e \ll (|\alpha/\eta|, |\beta/\eta|)$ we obtain:
\begin{eqnarray}
e_1 &\rightarrow& - \sqrt{\frac{1}{[\Lambda]_1}} \frac{\alpha}{\eta}  \ \ \ \ \ \phi_1 \rightarrow \frac{1}{\eta \tau_{e_1}} \nonumber \\ \nonumber \\
e_2 &\rightarrow& + \sqrt{\frac{1}{[\Lambda]_2}} \frac{\beta}{\eta}  \ \ \ \ \ \phi_2 \rightarrow \pi - \frac{1}{\eta \tau_{e_2}}.
\end{eqnarray}
where the involved quantities are given in terms of Keplerian orbital elements by equations (4), (19) and (20). Mathematically, $\Delta \phi \approx \pi$ arises from the fact that for all first-order resonances, $f_{\rm{res}}^{(1)} < 0$, while $f_{\rm{res}}^{(2)} > 0$. A physical consequence of this fact is that all stationary resonant planetary pairs will be apsidally anti-aligned.

\begin{figure*}
\label{Delta}
\includegraphics[width=1\textwidth]{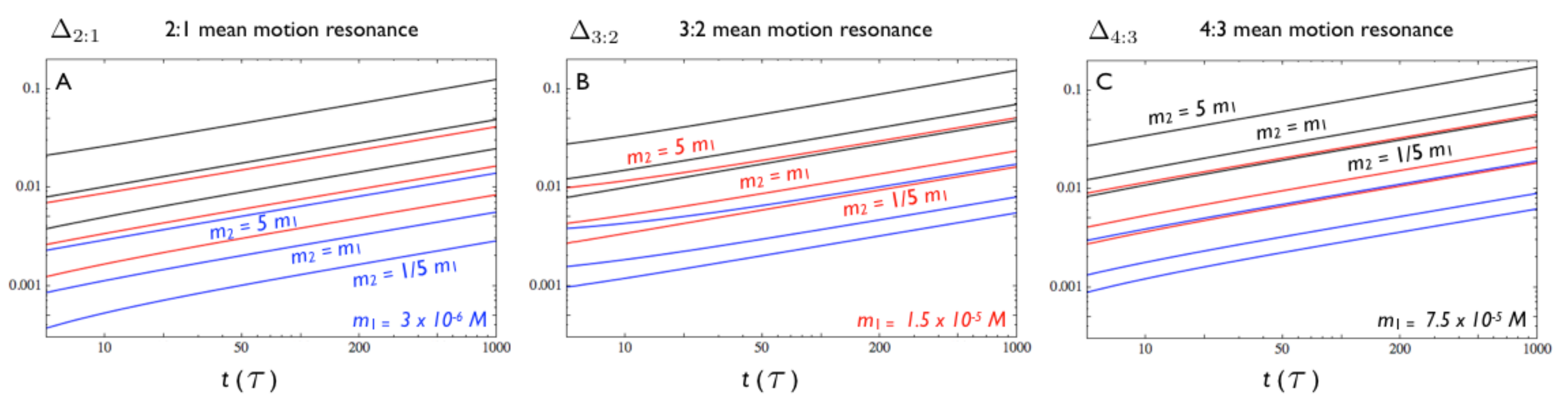}
\caption{The fractional extent of divergence away from nominal resonance, $\Delta$, as a function of the number of elapsed circularization timescales, $\tau$. The three panels correspond to the 2:1 (A), 3:2 (B) and 4:3 (C) mean motion resonances. The various plotted curves coincide with different mass ratios. Particularly, the blue, red and black curves are representative of $m_1 = 3 \times 10^{-6} M$, $m_1 = 1.5 \times 10^{-5} M$ and $m_1 = 7.5 \times 10^{-5} M$ respectively. As labeled in the Figure, for each choice of $m_1$, three choices of $m_2 = m_1/5$, $m_2 = m_1$, $m_2 = 5m_1$ are plotted, with the higher $m_2$ always corresponding to greater $\Delta$. For all calculations, we set $\tau_{e_1} = \tau_{e_2} = \tau$. Note that some systems can reach a fractional deviation from exact resonance of up to $\sim 20\%$, suggesting that dissipative divergence of resonant orbits is a viable mechanism for production of planet-pairs that reside significantly outside of nominal resonance.}
\end{figure*}

The above solution diverges as $\eta \to 0$ and gives positive values of $e_1, e_2$ only if $\eta<0$. This is because the stable equilibrium points of the resonance are always characterized by period ratios $n_1/n_2$ that are larger than the exact resonant value. This is a well-known fact for first order resonances (see for example Ch.9 of \citet{2002mcma.book.....M})\footnote{This is true only for small to moderate eccentricity values.}.  

The solution (29) also illustrates that, beyond the transient equilibration period, the eccentricity ratio remains constant for all time, since $\alpha$ and $\beta$ are strictly constant, while the actual eccentricity values depend only on $\eta$, i.e. on the proximity of the planets to exact resonance\footnote{Note that at the level of approximation which we have employed, the eccentric contribution to $\Psi$ can be neglected, since $\Phi \propto e^2$. Thus, in the definition of $\eta$ in (18) it can be safely assumed that $\Psi \simeq \Lambda$.}. Because we have restricted ourselves to only a linear treatment of eccentricity, this solution fails close to exact resonance, where equilibrium eccentricities can be quite large. However, this limitation only proves problematic in a rather narrow region of parameter space.

Thus far, we have only considered the relatively fast equilibration of orbital eccentricities and critical angles. Let us now turn our attention to the truly long-term evolution of the system and the associated change in the semi-major axes. There are two effects of importance. The simpler of the two effects is direct tidal damping of semi-major axes. To leading order in $e$ \citep{1966Icar....5..375G},
\begin{equation}
\label{dadttidal}
\left(\frac{d a}{dt}\right)_{\rm{tide}} = - 2e^2 \frac{a}{\tau_e}.
\end{equation}
Recall that the eccentricities converge onto quasi-fixed points. Thus, in terms of Poincar$\acute{\rm{e}}$ variables, the tidal decay of semi-major axes can be written as: 
\begin{eqnarray}
\label{dlambdadttide}
\left(\frac{d \Lambda_1}{dt}\right)_{\rm{tide}} = - 2  \frac{\Gamma_1}{\tau_{e_1}} \simeq -  \frac{1}{\tau_{e_1}} \frac{\alpha^2}{\eta^2} \nonumber \\
\left(\frac{d \Lambda_2}{dt}\right)_{\rm{tide}} = - 2  \frac{\Gamma_2}{\tau_{e_2}} \simeq -  \frac{1}{\tau_{e_2}} \frac{\beta^2}{\eta^2}.
\end{eqnarray}
For similar physical planetary parameters (including quality factors) and eccentricities, tidal evolution will cause orbits to diverge, since $\tau_{e_2}/ \tau_{e_1} \sim (k/k-1)^{10/3}$, although both semi major axes drift in the same direction (i.e. decay towards the central star).

The second, more subtle effect is the resonant divergence of the orbits, forced by eccentricity damping. As shown above, tidal decay of eccentricity causes the critical angles to collapse onto stable fixed points. However, these fixed points are slightly offset from the the actual foci. This offset results in a monotonic drift of the semi major axes in opposite directions. To understand this, let us return to our original formulation of the Hamiltonian. An application of Hamilton's equations to Hamiltonian (\ref{Hrespoincare}), evaluated on $e$ and $\phi$ given in (29), yields:
\begin{eqnarray}
\label{dlambdadtres}
\left( \frac{d \Lambda_1}{dt} \right)_{\rm{res}} &=& (1- k) (\frac{1}{ \tau_{e_1}} \frac{\alpha^2}{\eta^2} + \frac{1}{ \tau_{e_2}} \frac{\beta^2}{\eta^2} )   \nonumber \\
\left( \frac{d \Lambda_2}{dt} \right)_{\rm{res}} &=& k (\frac{1}{ \tau_{e_1}} \frac{\alpha^2}{\eta^2} + \frac{1}{ \tau_{e_2}} \frac{\beta^2}{\eta^2} ),
\end{eqnarray}
where we have made the small angle approximation: $\sin(\phi) \simeq \phi$. Note that the rate of change of the outer semi-major axis is positive definite, while that of the inner semi-major axis is negative definite. In other words eccentricity damping always results in the drift of the semi major axes in opposite directions, as anticipated above.

The long-term behavior of the resonance can be understood by combining equations (32), (18) and (29), to yield an equation of motion\footnote{Here, the direct tidal and resonant contributions to the evolution of the semi-major axes have been combined assuming that there are no indirect terms in the disturbing function i.e. the $\beta$'s in equations (\ref{dlambdadttide}) and (\ref{dlambdadtres}) are identical. This is true for all first-order resonant arguments, except $\phi = 2 \lambda_2 - \lambda_1 - \varpi_2$. In the exceptional case, proper account for the indirect terms must be taken (this is done in the calculation shown in Fig. 4).} for $\eta$:
\begin{equation}
\frac{d \eta}{dt} = -\frac{3 ([h]_1 (k-1) + [h]_2 k) ( k \alpha^2 \tau_{e_2} + (k-1) \beta^2 \tau_{e_1} )  }{\eta^2 \tau_{e_1} \tau_{e_2}}.
\end{equation}
This equation admits the solution
\begin{eqnarray}
\eta &=& (-1)^{2/3} \big{\{} \eta_0^3 - \frac{9 t }{\tau_{e_1} \tau_{e_2}} (k [h]_2 + (k-1) [h]_1 ) \nonumber \\
 &\times& ( k \alpha^2 \tau_{e_2} + (k-1) \beta^2 \tau_{e_1} )    \big{\}}^{1/3},
\end{eqnarray}
where $\eta_0 < 0$ is an initial condition, corresponding to the initial value of $\eta$ for a resonant equilibrium (which needs to be negative as shown in (29)) . Note that the solution (34) monotonically decreases in time, leading to an increase in the absolute value of $\eta$, i.e. an increase in the distance between the semi major axes of the planets relative to the Keplerian location of the resonance. The same $\eta \propto t^{1/3}$ dependence was observed in the simulations of \citet{2010MNRAS.405..573P}. Meanwhile, the resonant angles, $\phi$ will maintain a near-null libration width leading to quasi-constant eccentricity evolution.

Figure (4) presents an example of such evolution. In the case shown, two equal-mass ($m_1 = m_2 = 10^{-4} M$) planets are started out in exact 2:1 resonance with $a_1 = 0.05$ AU, $e_1 = e_2 = 0.01$, and randomly chosen angles. In this calculation, we have set $\tau_{e_1} = \tau_{e_2}$ and use this dissipation timescale as a unit of time (this is validated as a result of the adiabatic nature of the evolution). As above, each panel shows three separate calculations. Blue curves represent solutions obtained by numerically integrating the non-linear Hamiltonians (\ref{Hrespoincare}) and (\ref{Hkeplambda}) in presence of tidal dissipation (parameterized by equations (\ref{dedttide}) and (\ref{dadttidal})), red curves stem from the fully analytical framework presented in this section, while the gray curves result from an $N$-body simulation, where tidal and general relativistic interactions are accounted for directly \citep{2002ApJ...573..829M} and integrated using the Bulirsch-Stoer algorithm \citep{1992nrfa.book.....P}. As predicted by the theoretical arguments above, after a few ($\sim 5$) circularization timescales, the system collapses onto a fixed state where the critical angles approach their respective foci and the variations in eccentricities damp out. Once a quasi-stationary configuration is achieved, the orbits slowly diverge while the two resonant angles $\phi_1$ and $\phi_2$ remain in libration which means, strictly speaking, that the resonant configuration is maintained (although the separatrix associated with the resonance disappears at a certain $\eta$ - see \citet{2012arXiv1207.3171D, 1986sate.conf..159P}). 

Importantly, when dissipation is applied to a resonant pair, the outer orbit drifts outwards, gaining orbital energy. This behavior is in contrast with a naive application of standard tidal theory to the individual planets, where both planets are taken to drift inwards and facilitates a faster divergence of the orbits. 

As already mentioned above, the long-term evolution of the system is adiabatic: the characteristic timescale for significant orbital divergence greatly exceeds the resonant interaction timescale. Conveniently, this fact renders orbital divergence to be a scale-free process. In other-words, the fractional divergence away from exact resonance is not explicitly controlled by the actual semi-major axes or masses of the planets but rather by the mass-ratios ($m_1/m_2, m/M$) and the number of elapsed circularization timescales, $t/\tau$. Taking advantage of this, we have delineated the fractional extent of orbital divergence,
\begin{equation}
\Delta_{\rm{k:k-1}} = \frac{n_1/n_2 - k/(k-1)}{k/k-1}
\end{equation}
as a function of elapsed dimensionless time, $t/\tau$, for an array for planetary mass ratios. These results are demonstrated in Figure (5) where the three panels correspond to the 2:1 (A), 3:2 (B) and 4:3 (C) mean motion resonances. In the figure, blue curves correspond to $m_1 = 3 \times 10^{-6} M$, red curves to $m_1 = 1.5 \times 10^{-5} M$ and black curves to $m_1 = 7.5 \times 10^{-5} M$. For each color-coded choice of $m_1$, three choices of $m_2 = m_1/5$, $m_2 = m_1$, $m_2 = 5m_1$ are plotted, with the higher $m_2$ always corresponding to greater $\Delta$. Note that after $t/\tau \gtrsim 100$, the more massive examples presented in Figure (5), can reside more than $\sim 10\%$ away from nominal resonance. This points at the viability of creating the near-resonant overpopulation observed in the $Kepler$ sample by the the mechanism discussed here.

\section{Dissipative Dynamics of Multi-Resonant Planetary Systems}
There is considerable motivation to extend the above analysis to systems made of more than 2 planets, where each body is in resonance with all of its neighbors, as such systems appear to be common in nature. Perhaps the best-studied example of a multi-resonant system is the Galilean satellites, where both satellite pairs are locked in 2:1 mean motion resonances, leading to the libration of the Laplace argument. In the collection of confirmed extrasolar planets, examples of multi-resonant systems include the GL876 system - where the Laplace resonance is directly observed \citep{2010ApJ...719..890R}, the HD40307 \citep{2009A&A...493..639M} system  - which contains three planets that reside suspiciously close to a 4:2:1 period commensurability, as well as a few examples in the $Kepler$ data set. Furthermore, it has been shown that multi-resonant states can serve as good candidates for the initial condition of the solar system \citep{2007AJ....134.1790M, 2010ApJ...716.1323B}.

In this section, we shall extend our analytical theory of the long-term dissipative evolution of resonant configurations to systems that comprise more than 2 planets. As will be shown below, the dynamics of multi-resonant systems can be quite rich in diversity, so for simplicity, we shall work with a system consisting of three planets, keeping in mind that extension to a larger number of resonant objects can be accomplished.

As above, let us begin by writing out the full Hamiltonian. The Keplerian part reads:
\begin{equation}
\label{Hkep3body}
\mathcal{H}_{\rm{kep}} = - \frac{G^2 M^2 m_1^3}{2 \Lambda_1^2} - \frac{G^2 M^2 m_2^3}{2 \Lambda_2^2} - \frac{G^2 M^2 m_3^3}{2 \Lambda_3^2},
\end{equation}
while the resonant contribution is:
\begin{eqnarray}
\label{Hrespoincare2}
\mathcal{H}_{\rm{res}} &=& - \frac{G^2 M m_1 m_2^3}{[\Lambda]_2^2} ( f_{\rm{res}}^{(1,\rm{in})} \sqrt{\frac{2 \Gamma_1}{[\Lambda]_{1}}} \cos(\xi_1)  \nonumber \\
&+& f_{\rm{res}}^{(2,\rm{in})}  \sqrt{\frac{2 \Gamma_2}{[\Lambda]_{2}}} \cos(\xi_2^{\rm{in}} ) ) \nonumber \\
&-& \frac{G^2 M m_2 m_3^3}{[\Lambda]_3^2} ( f_{\rm{res}}^{(1,\rm{out})} \sqrt{\frac{2 \Gamma_2}{[\Lambda]_{2}}} \cos(\xi_2^{\rm{out}})  \nonumber \\
&+& f_{\rm{res}}^{(2,\rm{out})}  \sqrt{\frac{2 \Gamma_3}{[\Lambda]_{3}}} \cos(\xi_3) ),
\end{eqnarray}
where the four harmonics are:
\begin{eqnarray}
&\xi_1& = k^{\rm{in}} \lambda_2 - (k^{\rm{in}} - 1)  \lambda_1 + \gamma_1 \nonumber \\
&\xi_2^{\rm{in}}& = k^{\rm{in}} \lambda_2 - (k^{\rm{in}} - 1)  \lambda_1 + \gamma_2 \nonumber \\
&\xi_2^{\rm{out}}& = k^{\rm{out}} \lambda_3 - (k^{\rm{out}} - 1)  \lambda_2 + \gamma_2 \nonumber \\
&\xi_3& = k^{\rm{out}} \lambda_3 - (k^{\rm{out}} - 1)  \lambda_2 + \gamma_3
\end{eqnarray}
and the superscripts ``in" and ``out" refer to the resonances of the inner and outer pair of planets respectively. Before proceeding further, we note an important difference with the formalism developed in the previous section. In the two planet case, dissipation caused both critical angles to collapse onto their respective foci. Let us examine if similar behavior is possible in the three planet case. 

Suppose all four critical angles have evolved to a state where $d\xi/dt = 0$. In this case, simultaneous zero-amplitude libration of $d\xi_1/dt - d\xi_2^{\rm{in}}/dt = 0$ and $d\xi_2^{\rm{out}}/dt - d\xi_3/dt = 0$ implies that the apses of the system are locked i.e. $d\gamma_1 /dt = d\gamma_2 /dt = d\gamma_3 /dt = d\gamma_{\rm{sys}}/dt$. At the same time, expressing the mean longitude as $d\lambda/dt = n - d\gamma_{\rm{sys}}/dt$, the relationship $d\phi_2^{\rm{in}}/dt - d\phi_2^{\rm{out}}/dt = 0$ implies a strict correspondence among the semi-major axes: $- k^{\rm{out}} n_3 + (k^{\rm{in}}+k^{\rm{out}} -1 ) n_2 - (k^{\rm{in}} - 1) n_1 = 0$. A configuration that obeys this relationship is in (or close to) nominal resonance (e.g. the Galilean satellites). This means that away from nominal resonance, only three out of four critical angles can reside at their respective foci, while the remaining angle will circulate with the frequency
\begin{equation}
\label{xicirc}
d\xi_{\rm{circ}}/dt = - k^{\rm{out}} n_3 + (k^{\rm{in}}+k^{\rm{out}} -1 ) n_2 - (k^{\rm{in}} - 1) n_1.
\end{equation}

Naturally, if the system is far from nominal resonance, this circulation is comparatively fast, allowing us to drop (i.e. average over) the quickly varying harmonic and reduce the Hamiltonian (\ref{Hrespoincare2}) to a form that only contains three terms. This would further let us construct new action-angle coordinates, ensuring that the momenta conjugated to the three mean longitudes become constants of motion. However, identifying the circulating angle is not trivial a-priori, since the calculation inevitably depends on the planetary physical parameters, and in some cases can have non-linear dependence on initial conditions. Thus, unlike the two-planet problem described above, multi-resonant systems should be treated on a more case-by-case basis, as the construction of a suitable analytical theory for the long-term evolution depends on the properties of the system. Fortunately, as we already showed above, the timescale for the system to reach a quasi-stationary state is not much greater than the circularization timescale. So the initial transient period of system equilibration can be calculated numerically at a mild computational cost.

Due to the individual attention that multi-resonant planetary systems deserve, we shall leave the in-depth analysis of detected objects to follow-up papers and instead limit ourselves to an illustrative example of the long-term dynamical evolution of an equal-mass ($m_1 = m_2 = m_3 = 10^{-4} M$) planetary system in a 4:2:1 resonance. The aim of the calculation is largely to highlight the subtle differences between the evolution of a multi-resonant system and the results obtained for a single planetary pair in the previous sections. 

\begin{figure*}
\label{multires}
\includegraphics[width=1\textwidth]{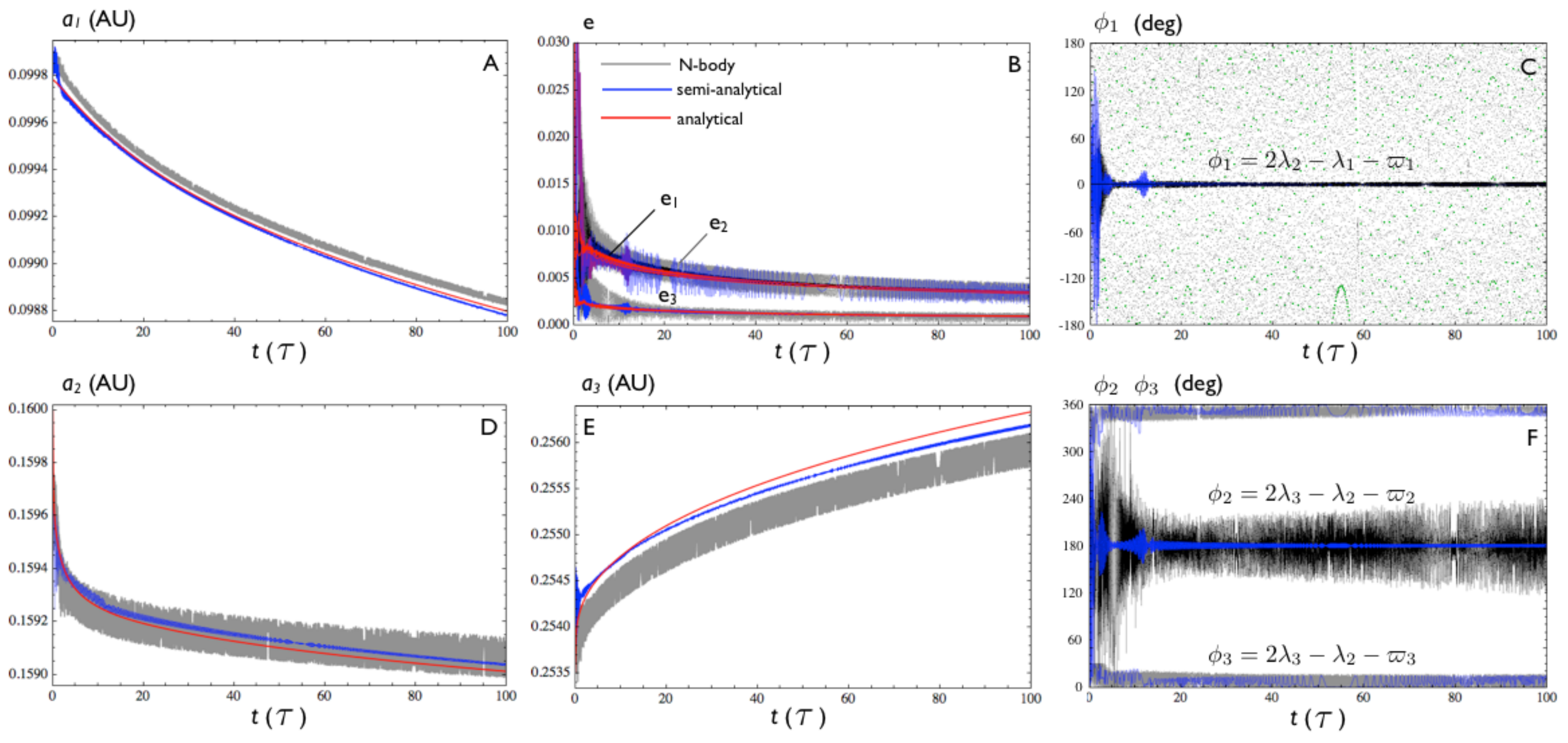}
\caption{Dissipative evolution of an equal-mass ($m_1 = m_2 = m_3 = 10^{-4}M$) planetary system in a 4:2:1 multi-resonant state over $t/\tau = 100$ circularization timescales. As in the two-planet case, the system settles onto a quasi-stationary state. However, the associated timescale is somewhat longer: $t \sim 10 \tau$. As discussed in the main text, only three of four critical angles can equilibrate, while the remaining angle is forced to circulate when far from nominal resonance. For the particular setup considered, as shown in panels C and F, the angles that tend to their respective foci are $\xi_1$, $\xi_2^{\rm{out}}$ and $\xi_3$.  Meanwhile, the gray (N-body) and green (semi-analytical) dots in panel C show the rapid circulation of $\xi_2^{\rm{in}}$. Panels A, D and E show the evolution of the planetary semi-major axes. In contrast to the two-planet calculation, here the drift of $a_2$ is inward rather than outward. Finally, the eccentricity evolution is shown in panel B. Although $e_1$ and $e_3$ settle onto quasi-stationary values, $e_2$ is significantly affected by the circulation of $\xi_2^{\rm{in}}$, never allowing the eccentricity to fully equilibrate. As before, the red curves were obtained analytically, while he blue curves were obtained by numerically integrating the equations of motion that arise from the Hamiltonians (\ref{Hkep3body}) and (\ref{Hrespoincare2}), augmented with a simple parameterization of tidal dissipation (i.e. equations (\ref{dedttide}) and (\ref{dadttidal})). The gray curves were computed numerically with a direct N-body simulation where dissipation has been taken into account. Note the excellent quantitative agreement between the theory and the numerics.}
\end{figure*}

With foresight, we begin with the construction of new canonically conjugated coordinates using the following generating function (intended for the system at hand):
\begin{eqnarray}
F_2 &=& \lambda_1 \Psi_1 + \lambda_2 \Psi_2 + \lambda_2 \Psi_3 + (k^{\rm{in}} \lambda_2 - (k^{\rm{in}} - 1)  \lambda_1 + \gamma_1)\Phi_1 \nonumber \\
&+& (k^{\rm{out}} \lambda_3 - (k^{\rm{out}} - 1)  \lambda_2 + \gamma_2)\Phi_2 \nonumber \\
&+& (k^{\rm{out}} \lambda_3 - (k^{\rm{out}} - 1)  \lambda_2 + \gamma_3)\Phi_3,
\end{eqnarray}
which yields the variables
\begin{eqnarray}
\Psi_1 &=& \Lambda_1 + (k^{\rm{in}} -1) \Phi_1 \ \ \ \ \ \ \ \ \ \ \ \ \  \ \ \ \  \ \ \ \ \ \ \ \ \ \  \psi_1 = \lambda_1 \nonumber \\
\Psi_2 &=& \Lambda_2 - k^{\rm{in}} \Phi_1 + (k^{\rm{out}}-1) (\Phi_2 + \Phi_3)  \ \ \ \  \  \psi_2 = \lambda_2 \nonumber \\
\Psi_3 &=& \Lambda_3 - k^{\rm{out}} (\Phi_2 + \Phi_3)  \ \ \ \ \ \ \ \ \ \ \ \ \ \ \ \ \ \ \ \ \ \ \ \   \psi_3 = \lambda_3 \nonumber \\
\Phi_1 &=& \Gamma_1 \ \ \ \ \ \ \phi_1 = k^{\rm{in}} \lambda_2 - (k^{\rm{in}} - 1)  \lambda_1 + \gamma_1   \nonumber \\
\Phi_2 &=& \Gamma_1 \ \ \ \ \ \ \phi_2 = k^{\rm{out}} \lambda_3 - (k^{\rm{out}} - 1)  \lambda_2 + \gamma_2   \nonumber \\
\Phi_3 &=& \Gamma_2 \ \ \ \ \ \ \phi_3 = k^{\rm{out}} \lambda_3 - (k^{\rm{out}} - 1)  \lambda_2 + \gamma_3.
\end{eqnarray}
This choice of variables is appropriate when the angle $\xi_2^{\rm{in}}$ is in circulation. Dropping this harmonic from the Hamiltonian renders $(\Psi_1, \Psi_2,
\Psi_3)$ constants of motion (if instead, the circulating angle had
been $\xi_2^{\rm{out}}$, the choice of $\Psi_2$ and $\phi_2$ would
have been made as in (13), identifying $k$ in (13) with $k^{\rm in}$,
and the angle $\xi_2^{\rm{out}}$ would have been dropped from the
Hamiltonian).

After some manipulation (as described in the previous sections), the Hamiltonian takes on a simple form:
\begin{eqnarray}
\label{Hsimple3body}
\mathcal{H} &=& \eta^{\rm{in}} \Phi_1 + \eta^{\rm{out}} \Phi_2 + \eta^{\rm{out}} \Phi_3 + \alpha^{\rm{in}} \sqrt{2 \Phi_1} \cos(\phi_1) \nonumber \\
&+&  \alpha^{\rm{out}} \sqrt{2 \Phi_2} \cos(\phi_2) +  \beta^{\rm{out}} \sqrt{2 \Phi_3} \cos(\phi_3),
\end{eqnarray}
where as before,
\begin{eqnarray}
&\eta^{\rm{in}}& = 3 ([h]_1 (k^{\rm{in}} - 1) \Psi_1 - [h]_2 k^{\rm{in}} \Psi_2 ) \nonumber \\
&\eta^{\rm{out}}& = 3 ([h]_2 (k^{\rm{out}} - 1) \Psi_2 - [h]_3 k^{\rm{out}} \Psi_3 )
\end{eqnarray}
are the proximities to exact resonance. The coefficient $\alpha^{\rm{in}}$ is given by equation (\ref{alpha}) and analogously, 
\begin{eqnarray}
\label{alphaout}
\alpha^{\rm{out}} = - \frac{G^2 M m_2 m_3^3}{[\Lambda]_3^2} \frac{ f_{\rm{res}}^{(1,\rm{out})}}{\sqrt{[\Lambda]_{2}}} \nonumber \\
\beta^{\rm{out}} = - \frac{G^2 M m_2 m_3^3}{[\Lambda]_3^2} \frac{ f_{\rm{res}}^{(2,\rm{out})}}{\sqrt{[\Lambda]_{3}}}.
\end{eqnarray}

As shown in the previous section, under dissipation the system will approach a quasi-stationary state. Once such a state is achieved, the corresponding fixed-point orbital parameters take on a familiar form:
\begin{eqnarray}
\label{threebodyfixedpoint}
e_1 &\rightarrow&  - \sqrt{\frac{1}{[\Lambda]_1}} \frac{\alpha^{\rm{in}}}{\eta^{\rm{in}}}  \ \ \ \ \ \ \phi_1 \rightarrow \frac{1}{\eta^{\rm{in}} \tau_{e_1}} \nonumber \\ \nonumber \\
e_2 &\rightarrow&  -\sqrt{\frac{1}{[\Lambda]_2}} \frac{\alpha^{\rm{out}}}{\eta^{\rm{out}}} \ \ \ \ \ \phi_2 \rightarrow \frac{1}{\eta^{\rm{out}} \tau_{e_2}} \nonumber \\ \nonumber \\
e_3 &\rightarrow&  +\sqrt{\frac{1}{[\Lambda]_3}} \frac{\beta^{\rm{out}}}{\eta^{\rm{out}}} \ \ \ \ \ \phi_3 \rightarrow \pi - \frac{1}{\eta^{\rm{out}} \tau_{e_3}}.
\end{eqnarray}
It is important to recall that we have dropped a quickly varying resonant term from the Hamiltonian when deriving these equations\footnote{Had the quickly varying harmonic been $\xi_2^{\rm{out}}$ instead of $\xi_2^{\rm{in}}$, the coefficients in front of terms containing $\Phi_2$ in (\ref{Hsimple3body}) would have been $\beta^{\rm in}$.  Equations (\ref{threebodyfixedpoint}) would then be modified accordingly.}. While the dropped harmonic will have little long-lasting effect, it will act to introduce high-frequency ``noise" into the solution, whose amplitude depends on the proximity of the system to exact three-body resonance. Thus, the equilibrium eccentricities and critical angles derived here are representative of average values.

Thus far, the behavior inferred from the above equations appears quite similar to the case of a single resonant pair described in the previous sections. However, an important difference surfaces when we consider the resonant drift of the semi-major axes:
\begin{eqnarray}
\label{dlambdadtresthreebody}
\left( \frac{d \Lambda_1}{dt} \right)_{\rm{res}} &=& \frac{1- k^{\rm{in}}}{ \tau_{e_1}} \frac{(\alpha^{\rm{in}})^2}{(\eta^{\rm{in}})^2}    \nonumber \\
\left( \frac{d \Lambda_2}{dt} \right)_{\rm{res}} &=& \frac{k^{\rm{in}}}{ \tau_{e_1}} \frac{(\alpha^{\rm{in}})^2}{(\eta^{\rm{in}})^2}  + (1- k^{\rm{out}})  \nonumber \\
&\times& \left(  \frac{1}{\tau_{e_2}} \frac{(\alpha^{\rm{out}})^2}{(\eta^{\rm{out}})^2} +\frac{1}{\tau_{e_3}} \frac{(\beta^{\rm{out}})^2}{(\eta^{\rm{out}})^2}  \right)  \nonumber \\
\left( \frac{d \Lambda_3}{dt} \right)_{\rm{res}} &=& k_{\rm{out}} \left(\frac{1}{\tau_{e_2}} \frac{(\alpha^{\rm{out}})^2}{(\eta^{\rm{out}})^2}  + \frac{1}{\tau_{e_3}} \frac{(\beta^{\rm{out}})^2}{(\eta^{\rm{out}})^2}  \right).
\end{eqnarray}
As in the two planet case, the drifts of the innermost and outermost planets are inward and outward respectively. The migration direction of the second planet, however, depends on the relative strengths of the inner and outer resonances, since the first term is positive definite while the second term is negative definite. Indeed, one could envision a set of system parameters (e.g. $m_3 \ll m_2,m_1$) where tidal dissipation leads to a divergence away from one set of resonances (increasing $|\eta^{\rm in}|$) and convergence onto another set of resonances (decreasing $|\eta^{\rm out}|$). In the context of such a scenario, conservation of the null phase-space area occupied by a quasi-stationary orbit will lead to eccentricity growth (this can be inferred from equations (\ref{threebodyfixedpoint})). At the same time, it is important to recall that the presented equations were derived as an expansion around nominal resonance location (which is assumed constant) and thus require dissipation in order to give rise to the corresponding drift of the semi-major axes. That is, one could in principle envision a scenario where only $\tau_{e_{1}}$ is finite, for which equations (\ref{dlambdadtresthreebody}) would predict a diverging inner pair and a stationary outer-most planet, inconsistent with resonant capture (and the associated drift of the nominal resonance location, $d[\Lambda]/dt$). However, as already pointed out above, the resonant harmonics are non-linearly coupled. Consequently, such a situation is atypical in practice, since dissipation on a single planet also results in damping of the other planet's eccentricities.

The application of the developed theory is demonstrated in figure (6). For the particular illustrative setup considered here, the angles $(\phi_1, \phi_2, \phi_3)$ attain a near-focal state within $t \sim 10 \tau$ while the dropped harmonic continues its circulation as expected. Although all three eccentricities decay monotonically as before, there is a clear qualitative difference in the behavior of $e_2$ compared to that of the two-planet case. In particular, $e_2$ never settles onto a fixed point, and is instead continuously driven by the circulation of $\xi_2^{\rm{in}}$, which contains $\gamma_2$, an angle conjugated to $\Gamma_2 \propto e_2^2$. Perhaps unsurprisingly, $e_1$ and $e_3$ are not strongly affected by this circulation. 

A more important distinction between the 2-planet and 3-planet evolutions is the direction of the second planet's drift. Namely, the combined effect of tidal dissipation and resonant interactions is now to drive the middle planet inward, whereas the evolution of $a_2$ was positive definite in the 2-planet case. All of this hints at the wide variety of possible outcomes and the dynamical richness of the multi-resonant interactions in presence of dissipative forces.

\section{Discussion}

The primary aim of this work has been to formulate a simple, physically intuitive analytical theory for the dissipative divergence of resonant orbits. We began with a purely conservative treatment of a single resonant pair and showed that at sufficiently low eccentricities and limited libration amplitudes, resonant dynamics can be treated with a linear, integrable approximation to the full resonant Hamiltonian. We then introduced simply parameterized tidal dissipation into the equations of motion and showed that the system tends to a quasi-stationary state over a few eccentricity circularization timescales. The collapse of the critical angles onto near-focal values in turn results in a divergent drift of the semi-major axes such that the outer orbit continually gains orbital energy while the inner planet's orbit decays. We subsequently showed how the developed formalism can be extended to multi-resonant systems. However, we have limited ourselves to a single illustrative example of the evolution of a system near a Laplace-like resonance, as we argued that the parameter space available to multi-resonant systems is quite large, rendering individual modeling more cost-effective.

Overall, our results point at the distinct possibility that the dynamical architectures of numerous detected systems, whose orbits seem to lie outside of resonance on the basis of the observed orbital periods, are a result of resonantly-aided dissipative divergence of the orbits \citep{2011CeMDA.111...83P}, and thus  comprise a number of important implications. First, the explanation we propose suggests that protoplanetary disks are indeed conducive to forming resonant planetary systems, whose long-term survival is assured \citep{2008A&A...482..677C}. In combination with precise quantitative modeling, this constraint can likely yield important new insights into understanding the physical structure and evolution of protoplanetary disks (e.g. weakly turbulent). 

Second, as shown in section 3, depending on the mass ratio and the
elapsed time, resonant orbits can evolve up to tens of percent away
from nominal resonance. If such extreme evolution is common, it is
possible that many planetary systems are actually in resonance even if their 
orbital periods are apparently not in commensurability. 
In particular, we expect the period-ratio
statistics of newly-formed planetary systems to cluster more clearly around resonant values than those of an evolved sample (see \citet{2012arXiv1202.6328F} for an in-depth discussion of the current data set).

Third, the fact that the time-dependence of the orbital divergence is
related to the tidal circularization timescale can be used to infer
from the observed period ratio how many circularization timescales a
given system has evolved through, if the age of the system is known.
Such information is vital for constraining unobservable parameters of
extra-solar planetary systems such as the planetary tidal quality
factor \citep{1966Icar....5..375G}, whose origin remains largely unexplained and is among the most
poorly constrained values in astrophysics. Although the above
arguments hinge on the observationally elusive characterization of the
physical planetary properties, we can certainly expect the data to
improve continuously over the coming years allowing for these
calculations to be executed, eventually.
\\

\textbf{Acknowledgments}  \\ 
We thank Kleomenis Tsiganis, Peter Goldreich and Greg Laughlin for numerous useful conversations. During the preparation of this paper, we have become aware that Lithwick \& Wu (2012, \textit{submitted}) arrived at similar arguments simultaneously and independently. K. Batygin acknowledges supported from NASA's NESSF graduate fellowship.

\end{document}